# Non-volatile ferroelectric memory effect in ultrathin $\alpha$-In$_2$Se$_3$

Siyuan Wan[1, #], Yue Li[1, #], Wei Li[2], Xiaoyu Mao[1], Chen Wang[1], Jiyu Dong[3], Anmin Nie[3], Jianyong Xiang[3, *], Zhongyuan Liu[3], Wenguang Zhu[1,2] & Hualing Zeng[1,2, *]

1. International Center for Quantum Design of Functional Materials (ICQD), Hefei National Laboratory for Physical Science at the Microscale, and Synergetic Innovation Center of Quantum Information and Quantum Physics, University of Science and Technology of China, Hefei, Anhui 230026, China

2. Key Laboratory of Strongly-Coupled Quantum Matter Physics, Chinese Academy of Sciences, Department of Physics, University of Science and Technology of China, Hefei, Anhui 230026, China

3. State Key Laboratory of Metastable Materials Science and Technology, Yanshan University, Qinhuangdao 066004, China

# Contribute equally to this work

* Corresponding author

**Recent experiments on layered $\alpha$-In$_2$Se$_3$ have confirmed its room-temperature ferroelectricity under ambient condition. This observation renders $\alpha$-In$_2$Se$_3$ an excellent platform for developing two-dimensional (2D) layered-material based electronics with nonvolatile functionality. In this letter, we demonstrate non-volatile memory effect in a hybrid 2D ferroelectric field effect transistor (FeFET) made of ultrathin $\alpha$-In$_2$Se$_3$ and graphene. The resistance of graphene channel in the FeFET is tunable and retentive due to the electrostatic doping, which stems from the electric polarization of the ferroelectric $\alpha$-In$_2$Se$_3$. The electronic logic bit can be represented and stored with different orientations of electric dipoles in the top-gate ferroelectric. The 2D FeFET can be randomly re-written over more than $10^5$ cycles without losing the non-volatility. Our** approach demonstrates a protype of re-writable non-volatile memory with ferroelectricity in van de Waals 2D materials.

**Introduction**

Featured with retentive and switchable electric dipoles, ferroelectrics are ideal candidates for building non-volatile memory with low-power consumption and ultrafast logic operation.[1-4] To date, a variety of ferroelectric based memories with different device designs have been realized, such as ferroelectric capacitors,[5, 6] ferroelectric tunnel junctions (FTJs),[7-10] and ferroelectric field-effect transistors (FeFETs).[11-17] Among these applications, capacitive ferroelectric memory functions in a destructive way, where the device needs to be re-written after every single readout. Besides, capacitors have limited integrated level subject to the detectable signal defined by their volume.[18] For FTJs, they suffer the critical size effect challenge as it is difficult for conventional ferroelectrics to maintain stable electric polarization in the electron quantum tunneling regime, normally down to a few unit cells thickness.[19-21] As a contrast, in FeFETs there is no requirement of ferroelectric film thickness and the data readout is non-destructive. All the merits origin from the three terminal device structure of field-effect transistor, in which logic bit reading and writing are separated in the semiconductor channel and the ferroelectric gate respectively.

Most conventional FeFETs make use of bulk perovskite ferroelectric as the device gate dielectric. It has been shown that the resistance of the conducting channel could be effectively modulated with the retention time at several days.[13, 14, 17] Recently, as the rise of two-dimensional (2D) material, enormous efforts focus on developing 2D FeFETs with the aim of further minimization of modern electronics. For example, graphene was coupled with bulk ferroelectrics, such as lead zirconate titanate (PZT) or organic ploy(vinylidene fluoride-co-trifluoroethylene) (PVDF-TrFE) thin film, to function as a hybrid quasi-2D FeFET.[22-27] In this type of FeFET, graphene was employed as the conducting channel and its conductance could be switched by the reversible electric polarizations of ferroelectric thin film. However, among all the reported FeFETs, genuine atomically thin 2D ferroelectric as the gate dielectric has rarely been reported. The ferroelectric films used in conventional FeFET are usually with film thickness at the order of several hundred nanometer.[25, 28] In contrast, incorporating ultrathin 2D ferroelectric will significantly scale down the dimensions of the device, which in turn increases the effective gate electric field. Thereby, the required writing or erasing voltage to flip the electric polarization could be further reduced, leading to a low-power consumption memory device.

In this letter, we report a 2D FeFET, which consists of van der Waals (vdW) interacted atomic layers only. We employ graphene and 2D ferroelectric α-$In_2Se_3$ as the conducting channel and electric gate respectively. Layered α-$In_2Se_3$ has been confirmed with the existence of room-temperature ferroelectricity under ambient condition.[29-32] Its ferroelectric polarization is robust even down to 2D limit as a result of the intrinsic inter-locking of in-plane and out-of-plane electric dipoles stemming from the lattice structure.[29, 33] This distinct feature helps the stabilization of the out-of-plane electric polarizations against the depolarization electric field in ultrathin α-$In_2Se_3$, which is technically more important for developing high-density memories.[29, 31] Here, as a demonstration of its potential application in information storage, we explore the nonvolatile ferroelectric memory effect in ultrathin α-$In_2Se_3$ with a 2D FeFET. In the device, the conductance of graphene was efficiently modulated by switching the polarization of ferroelectric α-$In_2Se_3$ ultra-thin layers. The transfer characteristic curve of the FeFET displays a butterfly-like trace, which is due to the ferroelectric hysteresis and the unique semimetal nature of graphene. By using heavily doped Si covered with $SiO_2$ as a reference gate, the ferroelectric polarization of α-$In_2Se_3$ under external electric field of $5\times10^5$ V/cm was estimated to be 0.92 μC/$cm^2$. Benefit from the long-term retention of ferroelectric dipoles in the top electric gate, the channel resistance was non-volatile, showing the signature of a memory device. The electronic logic bit could be encoded with different resistance state of the device, which corresponded to different orientations of electric dipoles. The 2D FeFET could be switched over more than $10^5$ cycles without losing the non-volatility. Our work demonstrates a practical application of 2D ferroelectrics and provides a quantitative and non-destructive method to study the ferroelectric states at nanoscale.

**Results and Discussions**

The three-dimensional schematic view and optical image of our FeFET device are depicted in Figure 1. The 2D FeFET was fabricated with vdW materials in a vertical stacking order of α-$In_2Se_3$/hexagonal boron nitride/graphene. Graphene and ferroelectric α-$In_2Se_3$ were used as conducting channel and top-gate dielectric respectively. Atomically thin hexagonal boron nitride (hBN), monolayer or bilayer, was chosen as the buffer and insulating layer. The introduction of hBN layer not only improved the device interface properties, such as suppression of ion diffusion and leakage current,[34, 35] but also enhanced the intrinsic conductivity of graphene by reducing the phonon and Coulomb scattering.[36, 37] Heavily doped silicon covered with 300 nm $SiO_2$ served as the substrate as well as the back gate. All the vdW materials used in the FeFET were prepared by mechanical exfoliation method.[38] Detailed device fabrication process can be found in Methods. The crystal structure of ferroelectric α-$In_2Se_3$ is illustrated in Figure 1a. In each monolayer α-$In_2Se_3$, there are five atomic layers following the Se-In-Se-In-Se layer sequence in the c axis. Its out-of-plane electric polarization stems from the dramatic difference of the interlayer spacings between the middle Se layer and its neighboring two In layers.[29] The room-temperature out-of-plane ferroelectricity of α-$In_2Se_3$ has been confirmed by recent experimental studies with piezo force microscopy (PFM) and second harmonic generation (SHG) spectroscopy.[30-32, 39, 40] Based on ferroelectric α-$In_2Se_3$ thin layers, we successfully fabricated five FeFET devices (details can be found in Supplementary Figure S1). All the FeFETs displayed the same electrical transport behavior. Figure 1b shows the optical image of one of the as-prepared FeFETs. The thickness of ferroelectric α-$In_2Se_3$ thin layers (~50nm, Figure 1c) was characterized by atomic force microscopy (AFM). We took the advantage of Raman spectrum to verify the crystal phase of α-In$_2$Se$_3$ used in the FeFET. As shown in Figure 1d, five characteristic peaks at 89 cm$^{-1}$, 104 cm$^{-1}$, 180 cm$^{-1}$, 196 cm$^{-1}$ and 266 cm$^{-1}$ were observed. These phonon modes were in good agreement with the results from previous studies on ferroelectric α-In$_2$Se$_3$,[30-32, 39] such as the softening of A$_1$ phonon mode at 104 cm$^{-1}$.

Figure 2 summarizes the electrical transport behavior of our 2D FeFET. The transfer characteristic of the device is nonlinear and hysteretic. When sweeping the ferroelectric gate voltage cyclically from negative to positive and back to negative, the resistance of the graphene channel could be efficiently modulated with large and stable hysteresis as shown in Figure 2a. The observed electrical hysteresis loop could be regarded as a reflection of the ferroelectric nature of α-In$_2$Se$_3$. By applying the top-gate voltage (V$_{TG}$), the electric dipoles in α-In$_2$Se$_3$ were electrically polarized in the out-of-plane direction, either upward or downward. The electric polarizations would significantly dope graphene with different types of carriers via the induced screening charges, leading to the shift of the Fermi level (E$_F$) of the graphene. Accordingly, the resistance of the FeFET device was modulated. When the Fermi level was tuned to the Dirac point, the conducting graphene channel reached its maximum resistance (R$_{Max}$) state. We found two R$_{Max}$ states at V$_{TG}$ = +1.4 V and V$_{TG}$ = 0 V respectively in the forward (from negative to positive) and backward (from positive to negative) gate voltage sweeping. The apparent difference of R$_{Max}$ on gate voltage was due to the requirement of coercive electric field (E$_c$) for the reversal of the ferroelectric domains in α-In$_2$Se$_3$. As a result, the resistance of our FeFET followed a butterfly-like dependence on gate voltage other than the ∧-shape as observed in most graphene field-effect-transistor (FET) with linear gate dielectric. By increasing the ferroelectric gate voltage, the hysteresis loop observed in the transfer characteristic curve was enlarged, due to the further electric polarization of ferroelectric domains in α-In$_2$Se$_3$. However, even under -2V or +3V gate bias, equivalent to an electrical field strength at the order of 10$^5$ V/cm in this device, the ferroelectric polarization in the top α-In$_2$Se$_3$ thin layers did not get saturated. Moreover, it should be pointed out that the observed hysteretic dependence of resistance on V$_{TG}$ allowed us to explore the orientations of electric dipoles and the electric polarization reversal in our FeFET. We marked four different points with I, II, III, and IV on the transfer characteristic curve in the top most panel of Figure 2a. Considering that the graphene used in this device was p-type (Supplementary Figure S2a), the corresponding configurations of electric dipoles at these four selected states, together with the different doping levels in graphene, were inferred as shown in Figure 2b. At point I and III, the bistable ferroelectric polarization in α-In$_2$Se$_3$ were established. The electric dipoles were well aligned to the external electric field, forming robust ferroelectric domains. With the upward electric polarization gating at point I, the Fermi level of graphene was lowered down below the Dirac point and it became a hole conductor. In contrast, at state III, it was n-type with electron doping. For the R$_{Max}$ state at point II or IV, it was the intermediated state during the electric polarization reversal process either from I to III or from III to I, where the electric dipoles were out of order. The exact correspondence of the electric dipole orientations to the resistance allowed us to encode the binary logic bit "0" and "1" in our FeFET. For example, we might utilize the bistable ferroelectric states at point I and III. More importantly, with the retention feature of these ferroelectric dipoles, the information could be stored in the device.

In previous electrical transport studies of graphene and other vdW materials based FET, charge traps at defects or polar adsorbates have been frequently reported to cause similar hysteretic effects.[41-45] Such kind of phenomenon was unstable and could be significantly suppressed by vacuum treatment, post-annealing, and substrate chemical modification.[46-48] Here, we emphasize that the hysteresis loop in the electrical measurements of the FeFET stems from the ferroelectric polarization reversal other than the charge traps at the interfaces or surfaces of α-In$_2$Se$_3$ thin layers in the following three aspects. First, the electrical measurements were carried out in vacuum with a relative slow voltage sweeping rate at 0.1 V/s. When different voltage sweeping rates were applied, there was no appreciable differences in the transfer characteristic curves of the FeFET devices as shown in Supplementary Figure S3. In addition, the estimated coercive electric field of α-In$_2$Se$_3$ thin layers from the FeFET was on the order of $10^5$ V/cm. This E$_c$ strength was in good agreement with that reported in previous work with PFM measurements.[39] Last but most importantly, the hysteresis loop observed in our FeFET is highly reproducible and stable. In the same device, the R$_{Max}$ states were fixed at constant gate voltages in many measuring cycles. The butterfly-like transfer characteristic curve could be repeated from device to device (details can be found in Supplementary Figure S1). These results ruled out the possibility of the observed electric hysteresis mainly originating from the bound charges at the interface or defects.

With the confirmed electrical writing/readout of the ferroelectric states in the FeFET, we aimed to quantitatively study the electric polarization in α-In$_2$Se$_3$ thin layers. To do this, we used the doping introduced by normal dielectric gating as a reference. In the device, the SiO$_2$/Si substrate was utilized as a second gate (or back gate). Employing both the top ferroelectric gate and back SiO$_2$/Si gate, we preformed dual-gating electrical measurements. Figure 3a shows the resistance of FeFET as a function of V$_{TG}$ under different back-gate voltages (V$_{BG}$). To make a better clarification, we unfold the electrical hysteresis loop by symmetrically expanding the x axis. First of all, the background resistance decreased gradually when the V$_{BG}$ changed from +40 V to -40 V, due to the contribution from the uncovered graphene region of the ferroelectric top gate (Figure 1b). [49, 50] The resistance of graphene as a function of the V$_{BG}$ were shown in Supplementary Figure S2b. Under each V$_{BG}$, the resistance was gated to the maximum twice in the back and forth V$_{TG}$ sweeping. Every R$_{Max}$ was marked with triangles (red and blue for backward and forward V$_{TG}$ sweeping respectively). The required V$_{TG}$ for R$_{Max}$ was shifted when varying the V$_{BG}$, indicating that the R$_{Max}$ was dependent on both V$_{TG}$ and V$_{BG}$. For reaching R$_{Max}$, the Fermi level of graphene was tuned to cross its neutral Dirac point, which meant that the carrier doping generated from the ferroelectric α-In$_2$Se$_3$ gating and the normal

SiO2/Si gating (density noted as n$_{TG}$ and n$_{BG}$ respectively) should compensate each other. Therefore, it is convenient to use an equivalent capacitor divider model (as shown in Figure 3b) to quantify the electric polarization (P) in ferroelectric α-In$_2$Se$_3$.[41,42] We use C$_{TG}$ and C$_{BG}$ to represent the equivalent capacitances of top and back gate respectively. To maintain the graphene being at its neutral charge point, the introduced doping follows

$$\text{n}_{TG} = \text{n}_{BG} + n_{intrisic} \rightarrow \text{C}_{TG} \cdot V_{TG} = \text{C}_{BG} \cdot V_{BG} + n_{intrinsic} \quad (1)$$

, where n$_{intrinsic}$ is the intrinsic carrier density of graphene. The capacitance of the back gate is calculated to be 11.5 nF/cm$^2$ with $\text{C}_{BG} = \varepsilon_0 \varepsilon_{SiO_2}/t$, where $\varepsilon_0$ is the vacuum permittivity, $\varepsilon_{SiO_2}$ is the dielectric constant of SiO$_2$ (~3.9), and $t$ is the thickness of SiO$_2$. As the electric polarization of top ferroelectric gate is as much as $\text{C}_{TG} \cdot V_{TG}$ in our FeFET, using equation (1) we deduce the P-E hysteresis loop as figured in Figure 3c. The maximum induced P of the ferroelectric α-In$_2$Se$_3$ thin layer in this device is estimated to be 0.92 μC/cm$^2$ under the external field strength of 5×10$^5$ V/cm applied in the out-of-plane direction.

In non-volatile memory devices, the building block is the recognizable and retentive physical quantities, where information can be stored and accurately readout. In Figure 2, the bistable ferroelectric states of the α-In$_2$Se$_3$, the upward and downward electric polarization (noted as $P_\uparrow$ and $P_\downarrow$), were proposed to be utilized. However, the corresponding resistances ($R_{P_\uparrow}$ and $R_{P_\downarrow}$) of these two states in the FeFET were almost indistinguishable. The $\frac{\Delta R}{R}$ (defined as $\left|\frac{R_{P_\uparrow}-R_{P_\downarrow}}{R_{P_\uparrow}+R_{P_\downarrow}}\right|$) was negligible at only 1%, which failed to provide good contrast for the readout of the binary logic bit. To overcome this issue, the symmetrical butterfly-like resistance dependence on V$_{TG}$ in the FeFET needs to be broken. One feasible way is to shift the Fermi level of graphene far away from its Dirac point prior to the ferroelectric doping, leading to that the electric polarization of α-In$_2$Se$_3$ will only result in monotonical resistance modulation. Therefore, the $\frac{\Delta R}{R}$ can be effectively enlarged. In the dual-gating electrical measurements, the $E_F$ of graphene was slightly tuned by the back-gate voltage. When varying $V_{BG}$, the symmetric hysteresis loop was distorted, which could be clearly found in Supplementary Figure S4a. The $\frac{\Delta R}{R}$ was dependent on $V_{BG}$ and reached 4% under the $V_{BG}$ at +40 V (Supplementary Figure S4b). To further improve this contrast for a better recognition of $P_\uparrow$ and $P_\downarrow$, we utilized heavily doped p-type graphene ($E_F$ at -0.2 eV, estimated by the Raman spectrum in Supplementary Figure S5) as the conducting channel in a new FeFET, named as p-FeFET for clarification. As shown in Figure 4a, the operation characteristic curve of the p-FeFET followed a single hysteresis loop, showing dramatic difference in resistance for $R_{P_\uparrow}$ and $R_{P_\downarrow}$ when electrically polarized at $\pm 3$ V. The $\frac{\Delta R}{R}$ was increased by one order, reaching 58.5%. This high contrast provided the basis for the binary logic bit encoding in our p-FeFET. With well resolved $R_{P_\uparrow}$ and $R_{P_\downarrow}$, we next tested the retention of the p-FeFET device. We firstly applied pulsed $\pm 3$ V gate voltages to make the logic bit ("1" or "0") writing of the device. After that, the electrical readout was performed from time to time with low reading bias (at 5 mV). Figure 4b showed the time dependence of the channel resistance. Both the resistances for $P_\uparrow$ and $P_\downarrow$ decayed slowly to constants but with opposite directions. After +3 V polarizing, the R decreased, while it increased with negative polarization. It should be pointed out that the residual resistances were still distinguishable in the time span of 1000 s. The observed evolution of the resistance on time in our p-FeFET was in accordance with the expectation of the remnant electric polarizations in ferroelectric α-In$_2$Se$_3$. It could be evidenced by the fact that the residual resistances for $P_\uparrow$ and $P_\downarrow$ in Figure 4b were in good agreement with that of the zero-field states on the hysteresis loop as indicated by the red arrows in Figure 4a. The existence of retentive and recognizable resistance states paved the way for the application of our p-FeFET in non-volatile data storage.

Another important issue for ferroelectric memory devices is the sustainability (or fatigue) of the switchable electric polarization. It has been shown that the ferroelectricity normally decreases or even vanishes after repeated polarization reversals.[51] This effect is understood as a result of the generation of trapping charges or atomic scale distortions during the writing/erasing processes. To test the fatigue of our 2D FeFET, we applied AC voltage in the triangle wave form from -2 V to +4 V at a repetition rate of 10 Hz (Supplementary Figure S6a) to reverse the electric polarization rapidly. After over $10^5$ cycles of repeated writing and erasing, the electrical hysteresis loop of the FeFET were found to be well preserving as shown in Supplementary Figure S6b. This observation evidences that the ferroelectricity in 2D α-In$_2$Se$_3$ is very robust against the perturbation of external electrical field, which lays the foundation for developing practical vdW materials based non-volatile memory device.

**Conclusions**

In summary, we have demonstrated the hysteretic ferroelectric gating and the non-volatile memory effect in the hybrid 2D FeFETs. The bistable states of the ferroelectric polarization in α-In$_2$Se$_3$ can be electrically controlled and non-destructively read out. With an independent linear dielectric gating as a reference, we have quantitively estimated the magniteude of the electric polarization of ultrathin α-In$_2$Se$_3$. Our results in this study indicate that the integration of 2D ferroelectric with other vdW materials is promising for future electronic applications in nonvolatile information storage.

**Methods**

*Sample preparations and device fabrications.* The bulk single crystal graphite and hBN used in this study were purchased from 2D Semiconductors, Inc.. α-In$_2$Se$_3$ thin layers were synthesized by chemical vapor deposition method. The ultrathin vdW materials, such as graphene and hBN, were prepared by the mechanical exfoliation method on to heavily doped n-type Si substrate with 300 nm $SiO_2$ on top and PDMS (Gel-Pak, WF-60-X4). The heterostructure were fabricated via all dry transfer technique.[38] Pre-selected ultrathin vdW layers were transferred from PDMS onto $SiO_2$/Si wafer under an optical microscope with the help of a home-made XYZ mechanical manipulator. The process of the stacking was shown in Supplementary Figure S7. The FET devices were fabricated by standard optical lithography, with the electrodes made by deposition of 10 nm titanium and 80 nm aluminum via e-beam evaporator.

*Topography measurements and Raman spectroscopy.* The topography of the devices was characterized by AFM (Bruker Dimension Icon) at contact mode. Raman spectra were taken by using Horiba micro-Raman system (Labram HR Evolution) with 633 nm laser excitation. The on-sample power of the excitation was 100 μW.

*Electric measurements.* Keysight B2900 source meter was used to measure the I-V characteristics of our FeFET devices. All the measurements were performed in a vacuum chamber with the pressure at $5\times10^{-3}$ Pa. In the gate voltage sweeping measurements, different dwell times of every test point were used to verify that the transfer characteristic curves were independent of sweeping rate. The sweeping rate of 0.1V/s was finally chosen.

**Acknowledgments**

This work was supported by the National Key Research and Development Program of China (Grant No.2017YFA0205004, 2017YFA0204904, and 2018YFA0306600), the National Natural Science Foundation of China (Grant No.11674295, 11674299, 11374273, 11634011, and 51732010), the Fundamental Research Funds for the Central Universities (Grant No. WK2340000082 and WK2060190084), Anhui Initiative in Quantum Information Technologies, the Strategic Priority Research Program of Chinese Academy of Sciences (Grant No. XDB30000000), and the China Government Youth 1000-Plan Talent Program. This work was partially carried out at the USTC Center for Micro and Nanoscale Research and Fabrication.

**Author Contributions**

H.Z. conceived the idea and supervised the research. Z. L and J. X prepared the samples. S.W., W.L. and Y.L. fabricated the devices. S.W. and Y.L. carried out the transport and Raman spectrum measurements of the hybrid FeFETs. S.W., Y.L. and H.Z. analyzed the data, wrote the paper, and all authors commented on the manuscript.

**Author Information**

The authors declare no competing financial interests. Correspondence and requests for materials should be addressed to Jianyong Xiang (jyxiang@ysu.edu.cn) and Hualing Zeng (hlzeng@ustc.edu.cn).

**Figures**

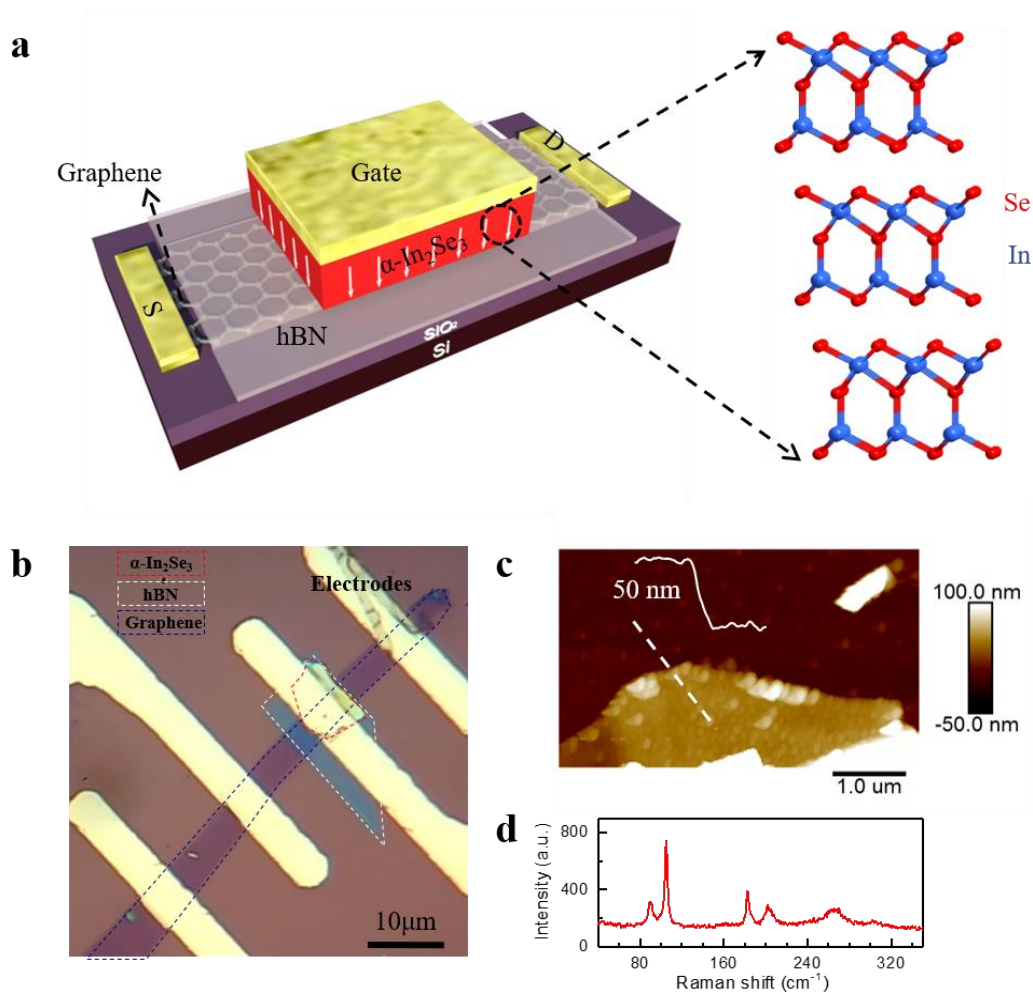

**Figure 1. Structure and optical characterization of the 2D FeFET. a) 3D schematic diagram of the FeFET. The FeFET is fabricated by vertically stacking graphene, hBN and α-In₂Se₃ thin layers in sequence. The white arrows indicate the direction of electric polarization. The zoomed area shows the crystal structure of ferroelectric α-In₂Se₃. b) Optical image of the FeFET device #1. The substrate is Si wafer with 300nm fused SiO₂ on top. The graphene, hBN, and ultrathin α-In₂Se₃ are indicated by black, white, and red dashed frames**

respectively. c) AFM topography and Raman spectrum of α-In₂Se₃ thin layers. The thickness of α-In₂Se₃ used in the device is 50nm. Five characteristic phonon modes of α-In₂Se₃ are observed in the Raman spectrum.

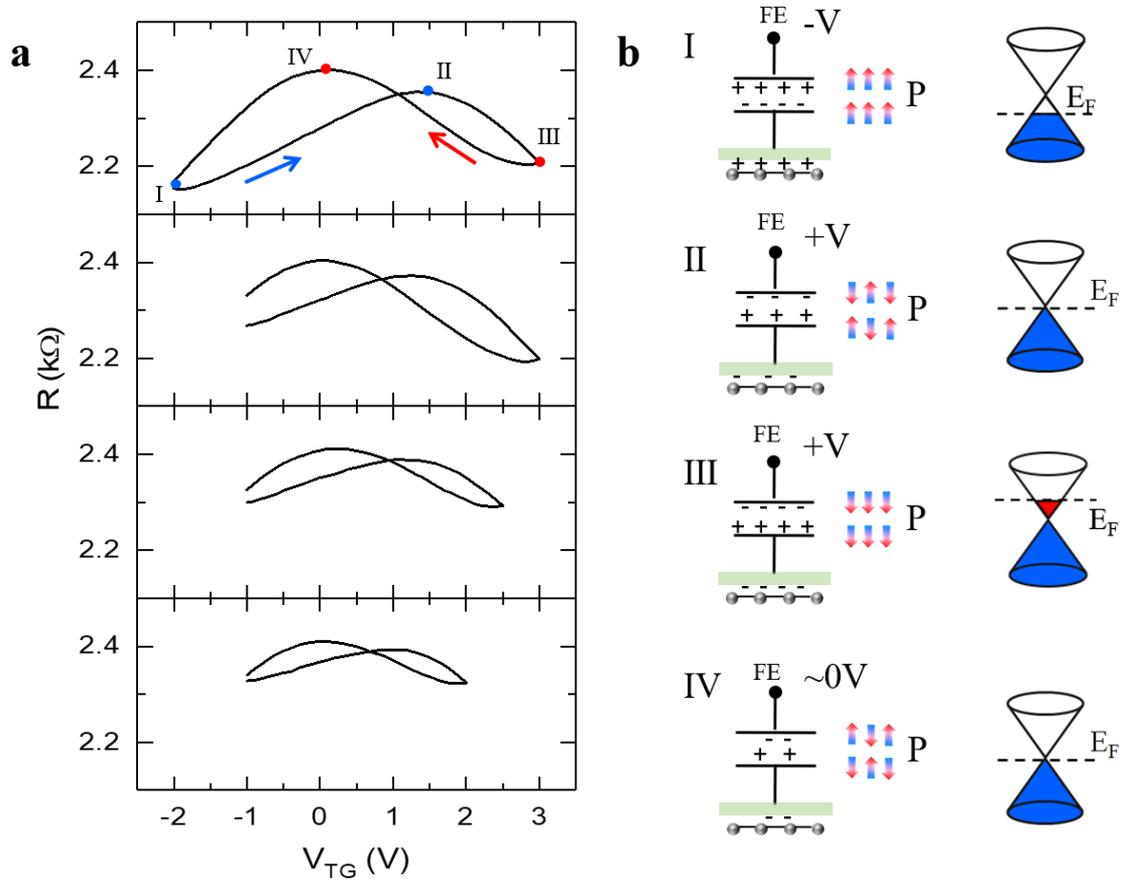

**Figure 2. Operation characteristics of the 2D FeFET. a)** The hysteretic ferroelectric gating in 2D α-In₂Se₃ based FeFET. The resistance follows a butterfly-like dependence on gate voltage. The electrical hysteresis loop can be enlarged by the range of the applied top gate voltage. **b)** Equivalent capacitor model of the 2D FeFET and the corresponding doping level in graphene. A capacitor is used to represent the top ferroelectric gate. The light green slab stands for the insulating hBN layer. The small color arrows represent the electric dipoles in α-In₂Se₃.

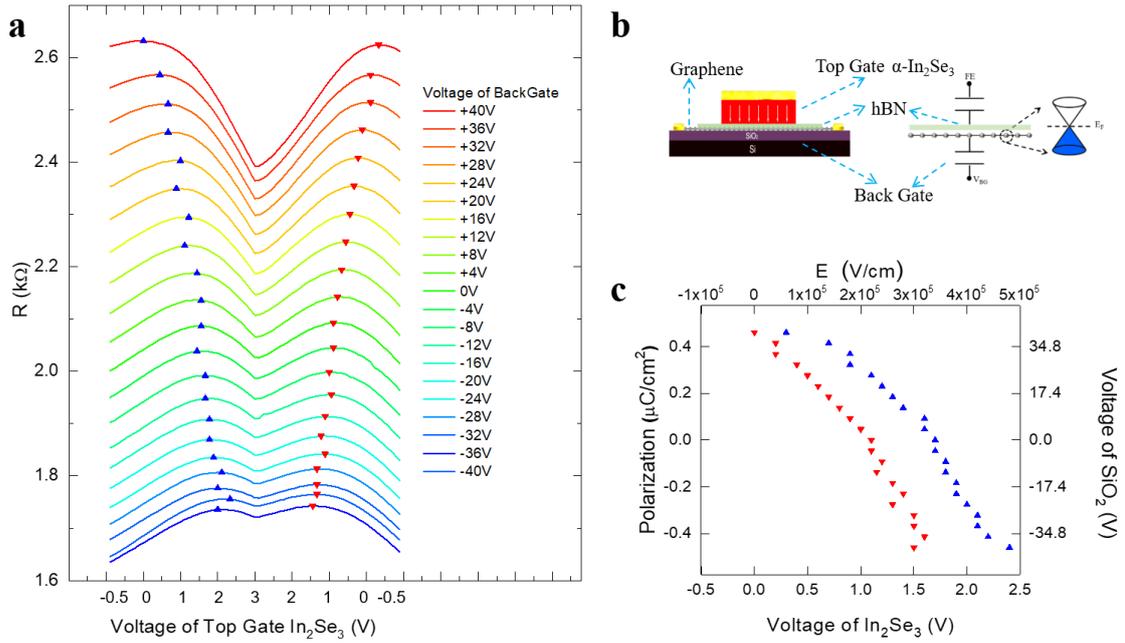

**Figure 3. Electric polarization of 2D ferroelectric α-In₂Se₃.** a) The channel resistance as the function of tog gate voltage under different $V_{BG}$. The x axis is symmetrically expanded for better clarification. The $R_{Max}$ states at the neutral Dirac points are marked by blue and red triangle respectively. b) Schematic of equivalent capacitor divider model. c) The P-E hysteresis loop of ferroelectric α-In₂Se₃ thin layers. Polarization is calculated by using the SiO₂/Si back gate as a reference without considering the intrinsic carrier density of graphene. The applied electric field strength is estimated by considering the film thickness of α-In₂Se₃ (50 nm).

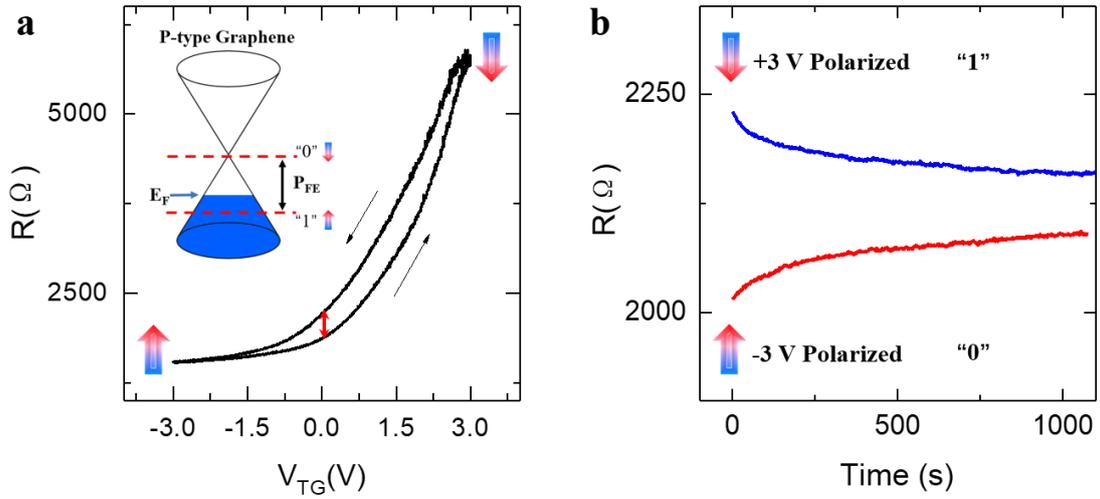

**Figure 4. Demonstration of non-volatile information storage in 2D FeFET. a) The hysteretic ferroelectric gating in p-FeFET. The upward and downward arrows represent the bistable ferroelectric polarizations in α-In$_2$Se$_3$. The small bidirectional red arrow in the hysteresis loop indicate the resistances states for the remnant electric polarizations. The inset shows the initial Fermi level of heavily doped p-type graphene. The red dashed lines indicate the final E$_F$ after the electric polarization of α-In$_2$Se$_3$ thin layers. b) The retention performance of the remnant ferroelectric polarization. The residual resistance is still recognizable after 1000s.**